# Quantum-Powered Optimization for Electric Vehicle Charging Infrastructure Deployment

Sakib, N. and Chen, X.

Department of Industrial Engineering

Southern Illinois University Edwardsville

**Abstract**

**The infrastructure development of electric vehicle charging stations (EVCS) is critical to the integration of electrical vehicles (EVs) into transportation systems, which requires significant investment and has long-term impact on the adoption of EVs. In this paper, a mathematical model is developed to identify the optimal placement of EVCS by utilizing a novel quantum annealing (QA) algorithm and quantum computation (QC). The objective of the optimization model is to determine the locations of EVCS that maximize their service quality for EV users. The model is validated using a real-world case study and solved using commercially available quantum computers from D-Wave. The case study shows that the QA algorithm can find the optimal placement of EVCS within seconds. The quality of the solutions obtained using QC is not sensitive to the shape or size of the area where EVCS are to be deployed.**

## I. Introduction

The demand for EVs is expected to increase significantly. This surge in demand highlights the need to construct public charging stations as an alternative to home-charging solutions, which in turn accelerate the adoption of EVs (Axsen & Kurani, 2013; Eberle & von Helmolt, 2010; Graham-Rowe *et al*., 2012; Lieven *et al*., 2011; Nie *et al*., 2016). The EVCSP problem is an optimization challenge that is gaining significant attention due to the rising number of EVs and the substantial cost of developing a public charging infrastructure for EVs. The EVCSP problem is a special case of the facility location optimization problems, which have been extensively studied in transportation and logistics (Aikens, 1985). The EVCSP problem is non-deterministic polynomial time (NP) hard, becoming computationally infeasible as the number of charging stations (CS) increases (Vazifeh *et al*., 2019). Solving the EVCSP problem using conventional computers with existing optimization algorithms is challenging.

The rise in real-world applications leveraging quantum computation (QC) has been notable since the inception of QC. Two robust and user-friendly QC tools are D-Wave's Ocean software development kit (SDK) and IBM's Qiskit. While the initial exploration of QC for real-life optimization applications began with simple tasks such as calculating prime numbers, researchers have delved into complex NP-hard problems (Au-Yeung *et al.*, 2022; Chatterjee *et al.*, 2023), which include the subset sum (Ghosh *et al.*, 2024), vertex covering (Chang *et al.*, 2009), graph coloring (Tabi *et al.*, 2020), and traveling salesman problem (TSP; Harwood *et al.*, 2021; Jiang & Chu, 2022). The QC hardware is still in its intermediate stage of development. The adoption of QC for optimization may be expedited by solving practical problems using the QC tools.

The primary motivation for this paper is to develop a new methodology for solving the EVCSP problem by leveraging the advanced QC optimization capabilities on the D-Wave Quantum Annealing (QA) platform. The EVCSP problem is solved using conventional computers and solvers, and the results are compared with those of QC optimization. Given the constraints of current quantum hardware, a quantum only approach to problem-solving is not yet feasible (Aaronson *et al.*, 2017). Nevertheless, this study demonstrates that QC can be used for optimization despite hardware limitations.

The rest of this article is organized as follows: Section II provides a concise overview of QC. Sections III and IV include an in-depth description of the proposed mathematical model for the EVCSP problem. Section V outlines the performance metrics used to validate both the proposed QC solvers and conventional solvers. The experimental results and discussions are presented in Section VI, while Section VII concludes the paper and outlines directions for future research.

## II. QC Optimization & EVCSP

The principles of quantum mechanics have led to the emergence of a new computing paradigm known as Quantum Computing (QC). This innovative approach harnesses the principles of superposition and entanglement, providing faster and more efficient solutions to complex industrial challenges that classical computers find difficult to solve. For instance, the capacity to rapidly determine the prime factors of large numbers on a quantum computer could undermine widely used encryption algorithms that rely on the almost infinite computation time of this problem. QC offers higher computing power, lower energy consumption, and exponentially higher speed compared to conventional computers. It achieves this by controlling the behavior of small physical objects, such as atoms, electrons, and photons. By manipulating properties such as superposition, entanglement, and interference, QC computers can store and process an enormous number of values simultaneously *(Kaswan et al., 2023)* using quantum bits (qubits), which are similar to bits in traditional computing.

Quantum superposition is a fundamental concept in quantum mechanics that allows a system to exist in multiple states simultaneously. In contrast to classical computing where a bit can be in either 1 or 0, a qubit is generally in a superposition state of basic state $|0\rangle$ and $|1\rangle$. This superposition nature of qubits allows for parallel computation and higher computing efficiency (Saravanamoorthi, 2023).

Equation (1) shows the mathematical formulation of the superposition of a qubit,

where $|\psi\rangle$ denotes the quantum state of the qubit, and $|0\rangle$ and $|1\rangle$ corresponds to the classical bits of 0 and 1, respectively. The coefficients $c_0$ and $c_1$ represent the probability of the qubit of being in states 0 and 1, respectively. Leveraging this superposition phenomenon, QC can perform massive quantum parallelism (Djordjevic, 2012). QC performs multiple parallel calculations in searching for solutions in a large solution space. As a result, quantum computers are more efficient and powerful than their classical analogs. The power of superposition becomes particularly evident in large-scale optimization problems, where qubits can explore multiple solutions simultaneously, leading to an exponential reduction in search time (Gondzio *et al.*, 2014). Searching for an optimal solution to an NP-Hard problem in classical computing is challenging. In contrast, a quantum computer can leverage parallel computing to swiftly explore near-optimal solutions in a fraction of a second. Figures 1.1 illustrates the quantum phenomena of superposition observed in quantum particles, where the blue circle indicates the pure state of 1 (spin up) and the red circle indicates the pure state of 0 (spin down). The superposition phenomena of this single qubit suggests that there is equal probability of collapsing into either of the pure states, 0 and 1.

$$|\psi\rangle = c_0|0\rangle + c_1|1\rangle \qquad (1)$$

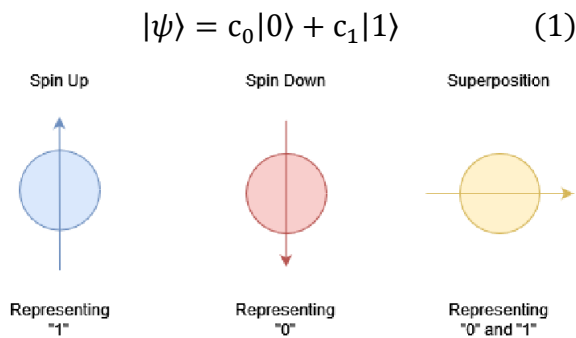

Fig 1.1: Superposition Phenomena of Qubits

Quantum entanglement is a phenomenon in quantum physics where two or more particles become connected and their quantum states are dependent on each other, even when they are separated by a large distance (Wang *et al.*, 2023). The correlation between qubits, known as entanglement, is non-local, allowing two qubits to become entangled regardless of the distance between them. Measurements of physical properties such as position, momentum, spin, and polarization performed on entangled particles can, in some cases, exhibit perfect correlation. For instance, if a pair of entangled particles is created with a known total spin of zero, and one particle is observed to have clockwise spin along a particular axis, the spin of the other particle, when measured along the same axis, will be found to be anticlockwise.

Qubits that initially exist independently without entanglement may be subjected to precise techniques, such as exposure to highly focused laser beams or radio waves, which result in the qubits' entanglement. Once entangled, the qubits enter an indeterminate state, where their properties are interdependent regardless of the physical distance between them. Even if the entangled qubits are separated by significant distances, they remain correlated or entangled. In QC, one of the entangled daughter particles (qubits) can be manipulated or subjected to operations, while measurements of their final states are taken. The entangled qubits lose their entanglement or become unentangled when their final states are observed or measured. Leveraging this phenomenon, quantum computing can perform calculations and serve as a powerful tool for various information processing tasks, including quantum teleportation, dense coding, quantum cryptography, and more.

Although QC is more efficient and powerful than classical computing, it is challenging to engineer and program QC due to de-coherence, noise, and highly fragile complex quantum states. While quantum phenomena like superposition and entanglement offer the promise of parallel computing, a significant challenge in achieving and maintaining these phenomena is de-coherence (Stilck França & García-Patrón, 2021). De-coherence occurs when the quantum behavior of qubits decays, often triggered by vibrations or fluctuations in temperature. These disturbances can abruptly disrupt the delicate quantum state of qubits, causing them to lose their superposition and potentially leading to computational errors. To mitigate this, qubits must be shielded from external influences using techniques such as supercooled refrigerators, insulators, or vacuum chambers to maintain the stability of quantum states. Industry leaders in quantum computing employ advanced technologies that maintain the qubits or quantum processing units (QPUs) in supercooled environments with temperatures as low as 15 millikelvin, colder than the depths of outer space. At these ultra-low temperatures, qubits can maintain stable superposition and entangled states, enabling quantum computing with minimal errors.

### A. QC Optimization Applications

Despite the impressive capabilities of today's supercomputers, many complex problems remain beyond the reach of conventional classical computers. QC represents a distinct and intricate approach to computation, characterized by two primary architectures: the gate model (also referred to as the circuit model) and quantum annealing (QA). The gate model implements algorithms using quantum gates, which function similarly to boolean gates in classical computers. On the other hand, in QA, the system begins in a low-energy state and progressively incorporates the variables of the problem to be solved. The gradual change increases the likelihood that the system will settle into a low-energy state that corresponds to an optimal solution for the problem. QA is implemented in D-Wave's publicly accessible quantum computers, e.g., the Advantage™, as a single quantum algorithm. This scalable approach to QC has enabled D-Wave to create QPUs with more than 5,000 qubits, which perform better than the state-of-the-art gate model.

Qiskit is an open-source SDK for working with QC. It offers tools for developing and manipulating quantum programs, allowing users to execute them on prototype quantum devices through the IBM Quantum Platform or on simulators installed on local computers. It adheres to the circuit model for universal quantum computing, making it compatible with any quantum hardware that operates based on this framework. The number of qubits in digital quantum computers, for example, quantum computers at IBM, is typically around 100, whereas the number of qubits in analog quantum computers, e.g., quantum computers at D-Wave, is around 2,000 qubits or more. With respect to the rate of development, the number of qubits in digital QC is increasing much faster and is expected to exceed the number of qubits in analog QC.

### B. QA

QA processors inherently yield low-energy solutions. Certain applications, such as optimization problems, seek the true minimum energy, while others, like probabilistic sampling problems, aim for high-quality low-energy samples. QA can help solve optimization problems by modeling them as energy minimization problems (Borowski *et al*., 2020). QA employs principles of quantum physics to identify low-energy states of a problem, thereby determining optimal or near-optimal solutions.

The qubits in the D-Wave quantum processing unit (QPU) represent the lowest energy states of the superconducting loops that constitute its architecture. These states exhibit a circulating current and generate a corresponding magnetic field. Similar to classical bits, a qubit can exist in either a 0 or 1 state. However, because the qubit is a quantum object, it can exist in a superposition of both the 0 and 1 states simultaneously. At the conclusion of the QA process, each qubit collapses from its superposition state into either 0 or 1, resulting in a classical state.

As qubits are added, systems become increasingly complex. For instance, two qubits can represent four possible states, while three qubits can represent eight states, thereby expanding the energy landscape significantly. Each additional qubit doubles the number of states defining the energy landscape, leading to exponential growth in the number of states as the number of qubits increases (Symons *et al*., 2023).

In summary, the QA process begins with a set of qubits in a superposition of 0 and 1, initially uncoupled. As QA progresses, couplers and biases are introduced, resulting in the entanglement of the qubits. At this stage, the system exists in an entangled state with multiple potential solutions. Upon completion of the annealing process, each qubit collapses into a classical state that signifies the minimum energy state of the problem or a state very close to it *(*Bunyk *et al., 2014)*.

## C. Optimal EVCSP

The EVCSP problem and its variations have been investigated using conventional computing methods. Recent research has focused on various innovative algorithms and methodologies to address the challenges posed by the increasing adoption of EVs. Among them, most studies have focused on electricity usage and its impact on the grid. A small number of studies are available that focus on optimal EVCSP. For instance, one study (Pratap *et al*., 2023) presents the gorilla troops optimization algorithm (GTOA) for determining the best placement, sizing, and power factor of distributed generation (DG) units alongside EVCS in radial power distribution networks (RPDNs). The study's goals are to minimize power loss, decrease voltage deviation, and improve voltage stability. Another study (Aljaidi *et al*., 2019) is a comprehensive survey of the widespread adoption of EVs and the critical role that the EVCSP plays in metropolitan areas. The paper outlines several challenges in the field, such as finding optimal CS locations in urban areas, determining CS charging demand, utilizing free public spaces for CS, increasing EV driver convenience, considering grid operator costs, preferred charging times for EV drivers, the distance between EVCS and the power grid, and external factors affecting access to CS.

Various algorithms and approaches, including exact methods and metaheuristics, e.g., genetic algorithm (GA) and particle swarm optimization (PSO), have been investigated to optimize EVCSP. Additionally, multi-objective optimization has been explored using different solution approaches to address multiple goals simultaneously (Lazari & Chassiakos, 2023). A few articles have examined the EVCSP problem from the perspective of minimizing the total travel distance to frequently visited CS and maximizing the distance between a new CS and other existing CS. Minimizing waiting time and idle time at EVCS has not been explored in these studies.

To the best of the authors' knowledge, limited research addresses the EVCSP problem using the newly available optimization capability of QC. Talebpour (2023) explored the possibility of solving EVCSP problem using Grover's quantum search algorithm. Chandra *et al.* (2021) used a hybrid meta-heuristic algorithm of GA and QA to solve the EVCSP problem. Hu *et al.* (2024) considered the environmental and economic factors in solving the EVCSP problem using an improved quantum genetic algorithm (QGA). While QC optimization appears promising in solving the EVCSP problem, the main challenge is to find an optimal solution given the probabilistic nature of qubits.

### III. Proposed QC Algorithm for EVCSP

Quantum-classical hybrid computing integrates both classical and quantum resources to solve complex problems, leveraging the strengths of each. This approach allows developers to benefit from current quantum capabilities while positioning for future advancements. As quantum processors grow in size, they offer unmatched performance for specific tasks, such as solving hard optimization problems. D-Wave Hybrid, for instance, provides a Python framework for building flexible workflows, enabling parallel use of quantum and classical resources. Hybrid workflows involve using classical methods to isolate the core problem or breaking large problems into smaller, quantum-solvable parts for optimal solutions.

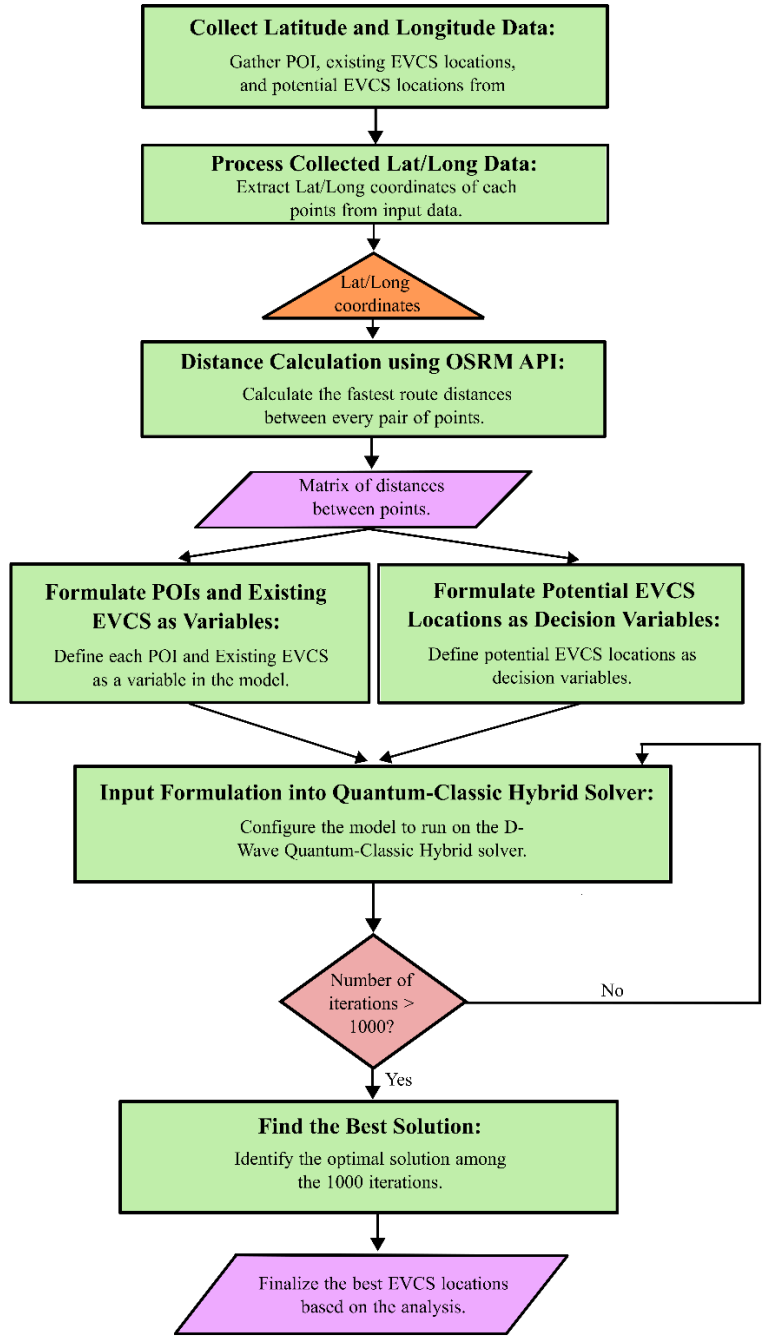

Fig 3.1: Workflow for Identifying Optimal EVCS Locations Using Quantum-Classic Hybrid Solver.

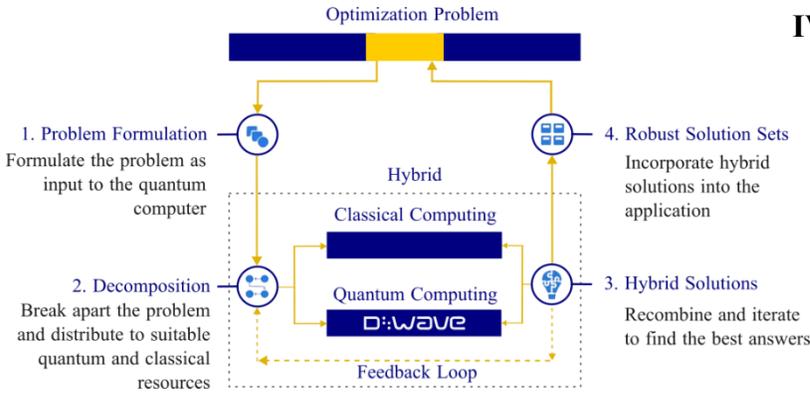

Fig 3.2: Working principles of Hybrid Quantum-Classical Computing on D-Wave

In Figure 3.1, we present the flowchart of our proposed model. Initially, we gathered comprehensive data for the selected city used in our case study from OpenStreetMap (OSM). Next, we extracted the latitude and longitude coordinates for various Points of Interest (POIs), including hotels, parking lots, parks, supermarkets, and existing EVCS locations. The latitude and longitude of these points are then used as input and fed into the OSRM API to calculate the fastest route distance between them. Most previous research in this domain has utilized the Haversine distance, which is less accurate compared to the OSRM fastest route distance. We obtain a distance matrix from the OSRM function, which is then used to formulate the objective function. POIs, hotels, and supermarkets are treated as variables, while potential EVCS locations are defined as decision variables. In the next stage, we solve the optimization model using Quantum-Classical hybrid solvers from D-Wave, running the solver for 1,000 iterations. The best solution is then selected from these iterations, determining the optimal EVCS locations.

## IV. Model Formulation

The EVCSP problem encompasses various scenarios. The focus of this study is towns and small cities in the U.S. The goal is to strategically place new CS to ensure convenience for all POIs in the area of interest. For instance, if the POIs represent shops along a main street, the optimal location may be a central point on the street that has a minimum average distance to all POIs. In the mathematical model, the average distance from a potential new CS to all POIs is used. In addition, a new CS should not be close to existing ones to maximize coverage and utilization. Fig. 4.1 illustrates the EVCS whose locations are randomly chosen. In Fig. 4.1, "P" denotes the location of POIs whereas the red circles represent EVCS.

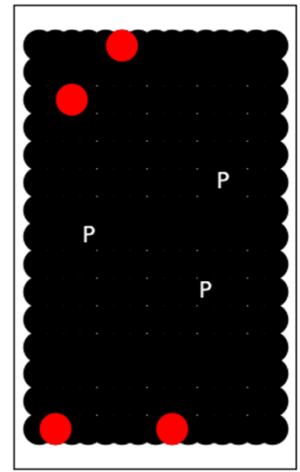

Fig 4.1: EVCS with Randomly Selected Locations.

The EVCSP problem is a specific type of general facility location problem that are NP-hard. The mathematical model is a mixed-integer linear programming (MILP) model with three objective functions. An experimental evaluation is conducted first with randomly generated small-size synthetic instances using exact solvers and generic

MILP solvers. Then the model and methods are validated using real-world data for the City of Edwardsville, Illinois, U.S.

**Nomenclature**

| Parameter | Description | Unit |
|---|---|---|
| $E$ | Total number of CS locations | |
| $P$ | Total number of POIs | |
| $CS$ | Total number of locations for new CS | |
| $d_{ij}$ | Distance between a POI ($i$) and a new CS ($j$) | KMs |
| $e_{ij}$ | Distance between an existing CS ($i$) and a new CS ($j$). | KMs |
| $q_{ij}$ | Distance between a new CS ($i$) and a new CS ($j$) | KMs |
| $x_j$ | Binary decision variable: $x_j = 1$ if the location $j$ is selected, $x_j = 0$ otherwise | |
| $\gamma_1$ | Lagrange multiplier for objective function 1 | |
| $\gamma_2$ | Lagrange multiplier for objective function 2 | |
| $\gamma_3$ | Lagrange multiplier for objective function 3 | |

Equation (2) is an objective function that minimizes the average distance between potential CS and POIs. The average distance is calculated by dividing the total distance by the total number of POIs. The average distance is multiplied by a Lagrange multiplier for quantum optimization.

$$Min\ Z_1 = \frac{\sum_{i=1}^{P} \sum_{j=1}^{E} d_{ij} \times x_j}{p} \times \gamma_1 \quad (2)$$

Equation (3) maximizes the distance between existing CS and potentials CS locations. The total distance is divided by total number of existing CS to calculate the average distance. The average distance is multiplied by a Lagrange multiplier for quantum optimization.

$$Min\ Z_2 = -\frac{\sum_{i=1}^{X} \sum_{j=1}^{E} e_{ij} \times x_j}{c} \times \gamma_2 \quad (3)$$

Equation (4) maximizes the distance between two potential CS locations. Similar to Equations (3), this objective function is multiplied by $-1$. Equation (5) ensures a fixed number ($CS$) of new CS are placed, where $x_j$ is a binary decision variable.

$$Min\ Z_3 = -\sum_{i=1}^{E} \sum_{j=1}^{E} q_{ij} \times x_j \times \gamma_3 \quad (4)$$

$$\sum_{j=1}^{E} x_j = CS \quad (5)$$

The Lagrange multipliers are tunable parameters that are used to control the objective functions. The values of the Lagrange multipliers are selected based on previous studies: $\gamma_1 = E \times 4$, $\gamma_2 = \frac{E}{3}$, and $\gamma_3 = E \times 1.7$.

## V. Case Studies

Two case studies, one with a random node generator and another with real-life data for the City of Edwardsville, are investigated. In the first case study, a 2D grid with a height and width of 20 kilometers, five POIs, and seven existing CS, is used. There are 148 potential CS locations from which four new CS locations are chosen. Fig. 5.1 shows the solution to the optimization model in the first case study. The red circles indicate the existing CS and "P" indicates the POIs. The blue circles indicate the new CS. The D-Wave QA solvers can solve the optimization model and determine the locations of the four new CS.

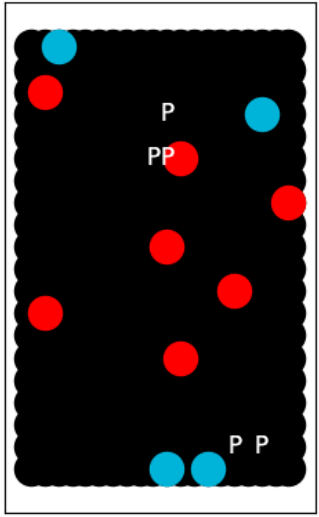

Fig 5.1: EVCS Locations in the Case Study.

In the second case study, data for the City of Edwardsville, Illinois are used to validate the optimization model. All data for POIs, existing CS, and potential CS are acquired from open-source Open Street Map (OSM) platform. Fig. 5.2 shows two existing EVCS, which cannot meet the demand of EV users in this region. Fig. 5.3 shows the POIs in Edwardsville that include supermarkets, hotels, and restaurants. The red and cyan points are hotels and supermarkets respectively. Orange points are restaurants. For this study, there are 10 supermarkets, three hotels, and 41 restaurants.

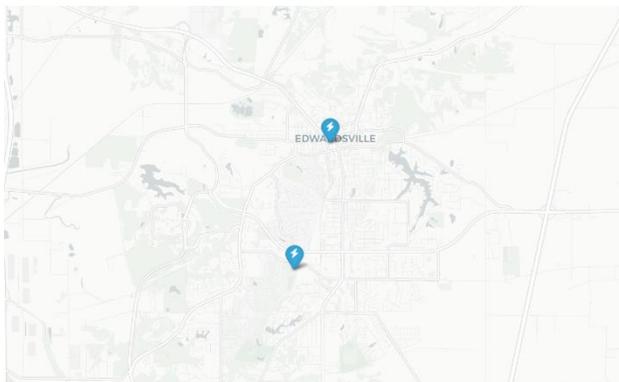

Fig 5.2: Existing EVCS in Edwardsville.

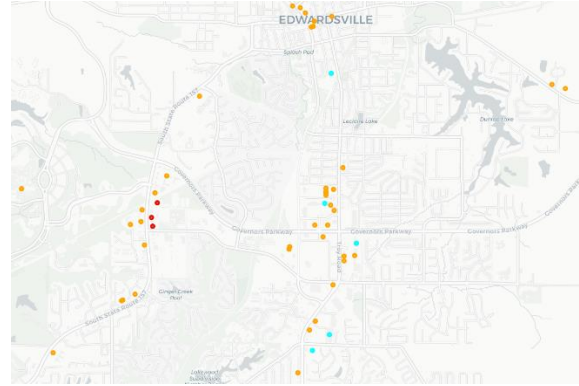

Fig 5.3: POIs in Edwardsville.

Parking spaces, public parks, and gas stations in Edwardsville are potential locations for new EVCS. Fig. 5.4 shows the potential locations in which the green points are public parking spaces, blue points are public parks, and magenta points are gas stations. There are 389 parking spaces, 30 parks, and 11 gas stations.

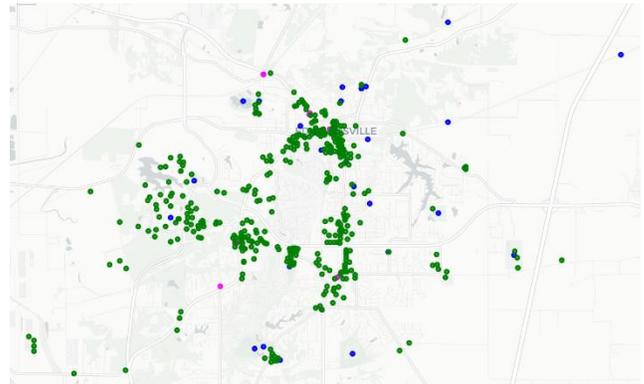

Fig 5.4: Potential CS Locations.

Table 5.1 summaries POIs and potential EVCS locations.

Table 5.1: POIs and EVCS Locations.

| Type | Name | Number of Locations |
|---|---|---|
| POIs | Hotels | 3 |
|  | Restaurants | 41 |
|  | Supermarkets | 10 |
| Existing EVCS |  | 2 |
|  | Parking Spaces | 389 |

| Potential EVCS Locations | Parks | 30 |
|---|---|---|
| | Gas Stations | 11 |

## VI. Result & Discussion

The hybrid quantum solver provides near optimal solutions due to quantum computers' heuristic nature. Fig. 6.1 shows the output from the solver. Exactly four EVCS are located based on the input. The average distances between POIs and these EVCS locations are 0.078, 0.030, 0.024, 0.027 Km. The average distances between the two existing CS and these four EVCS locations are 0.09, 0.008, 0.038, 0.039 Km. The average distance among these four EVCS locations is 0.337 Km.

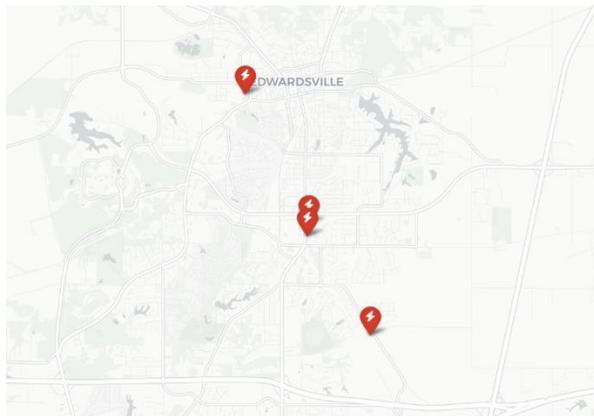

Fig 6.1: Selected EVCS Locations.

This optimization problem consists of 389 binary variables and the hybrid quantum solver was able to find the near optimal solutions in 2.997 seconds with a QPU access time of 0.080 seconds. The optimal locations are identified after multiple runs of the model to account for variabilities in results due to the probabilistic nature of QA. Different values for Lagrange multiplier have been tested and analyzed.

## VII. Conclusion

The EVCSP problem is a combinatorial optimization problem that aims to place the EVCS to meet multiple objectives. The most common objective is to minimize the average distance between POIs and potential CS locations. Previous research suggested that finding an exact solution to the EVCSP problem was not practical. QC is a novel approach that searches for the optimal solutions using parallel computing. In this study, a novel Hybrid Quantum Classical algorithm is used to find near optimal solutions to the EVCSP problem using the D-Wave hybrid quantum solvers. A real-life case study is analyzed using data for the City of Edwardsville, Illinois, U.S.